\documentclass[%
 aip,
 jmp,%
 amsmath,amssymb,
 reprint,%
]{revtex4-1}

\usepackage{graphicx}
\usepackage{dcolumn}
\usepackage{bm}

\begin{document}

\preprint{AIP/123-QED}

\title[Nonlinear Waves and Coherent Structures]{Nonlinear Waves and Coherent Structures in the Quantum Single-Wave Model\footnote{Error!}}

\author{Stephan I. Tzenov}

\affiliation{Department of Physics, Lancaster University, Lancaster LA1 4YB, United Kingdom}

\altaffiliation{The Cockcroft Institute, Keckwick Lane, Daresbury, WA4 4AD, United Kingdom.}


\email{s.tzenov@lancs.ac.uk}


\author{Kiril B. Marinov}

\affiliation{ASTeC, STFC Daresbury Laboratory, Keckwick Lane, Daresbury, WA4 4AD, United Kingdom.}

\altaffiliation{The Cockcroft Institute, Keckwick Lane, Daresbury, WA4 4AD, United Kingdom.}


\date{\today}

\begin{abstract}
Starting from the von Neumann-Maxwell equations for the Wigner quasi-probability distribution and for the self-consistent electric field, the quantum analog of the classical single-wave model has been derived. The linear stability of the quantum single-wave model has been studied, and periodic in time patterns have been found both analytically and numerically. In addition, some features of quantum chaos have been detected in the unstable region in parameter space. Further, a class of standing-wave solutions of the quantum single-wave model has also been found, which have been observed to behave as stable solitary-wave structures. The analytical results have been finally compared to the exact system dynamics obtained by solving the corresponding equations in Schrodinger representation numerically.
\end{abstract}

\maketitle

%

\section{\label{sec:intro}Introduction}

Conventional plasma physics describes dynamical and equilibrium processes in systems characterized by high temperatures and relatively low densities, at which quantum effects have practically no influence. Recent technological progress on miniaturized semiconductor devices and nanoscale structures has stimulated the interest in potential applications of plasma physics, where the quantum nature of the constituting objects plays an essential role.

Both fusion and space plasmas are characterized by states of high temperatures and low densities, for which quantum effects are negligible and are usually not taken into account. The classical example, where plasma and quantum effects mutually coexist is the electron gas in metals at room temperature. The electron population in ordinary metals is globally neutralized by the lattice ions and therefore can be regarded as a classical example of a plasma system. However, its basic properties are mainly determined by the presence of the regular ion lattice, so that typical plasma effects are only correction of higher order. In recent years, substantial progress has been made in preparation and manipulation of metallic nanostructures consisting of a small number of atoms (typically $10 - 10^5$) \cite{Calvayrac}. Since an underlying ionic lattice for such structures does not exist, the electron dynamics is dominated by plasma effects, at least for systems with sufficiently large spatial dimensions. This makes metallic nanostructures the ideal object to study equilibrium and evolution properties of quantum plasmas.

Additional important applications of quantum plasmas arise from semiconductor physics and astrophysics \cite{Kluksdahl,Markowich,Shapiro,Lai,Kastor}. The basic factor dictating the necessity of utilizing quantum mechanical description in semiconductors is the great degree of miniaturization of today's electronic components to the extent that the de Broglie wavelength of the charge carriers can be comparable to the spatial variation of the doping profiles. In some astrophysical objects under extreme conditions of temperature and density, such as white dwarf stars, the
density can become some ten order of magnitudes larger than that of ordinary solids. Because of such large densities, a white dwarf can be as hot as a fusion plasma ($10^8$ K), but still behave as a quantum-mechanical object.

When quantum effects start playing a significant role, the well-known picture from conventional plasma physics becomes even more complicated, because an additional length scale has to be introduced. This is the de Broglie wavelength of a charged particle, $\lambda_B = \hbar / (m v_T)$, where $v_T$ is its thermal velocity. The latter represents approximately the spatial extension of the particle wave function and obviously the larger it is, the more significant quantum effects are. It is also clear that quantum behavior will be reached more easily for the electrons than for the ions, due to the large mass difference. Indeed, in all practical situations, even in the most extreme ones, the ion dynamics remains always classical, while only the electrons require quantum-mechanical description.

Another important area uncovering broad vistas for applications of the quantum plasma dynamics approach is the free electron laser (FEL). If the photon momentum recoil is greater than, or of the order of the momentum spread in the electron beam traversing the undulator, quantum effects contribute substantially and cannot be neglected \cite{Serbeto}.

Due to the tremendous number of potential applications ranging from recent high technology developments to high energy physics and astrophysics, there has been a growing interest in studying new aspects of dense quantum plasmas \cite{Shukla}.

The purpose of the present paper is to extend the classical single-wave model \cite{DelCastillo,MarTze} to the quantum case. Following del-Castillo-Negrete \cite{DelCastillo} we give in the next Section a brief sketch of the derivation of the quantum single-wave model. In Sections \ref{sec:schrodpic} and \ref{sec:perturbative} the perturbation solution of the quantum single-wave model in Schrodinger picture for a mixed state with different energies has been constructed, and in Section \ref{sec:amplitude} the amplitude equation describing the nonlinear wave interaction between eigenmodes has been obtained. Further, in Section \ref{sec:standing} a class of solutions of the quantum single-wave model in the form of nonlinear standing waves has been found, much resembling solitary wave structures. In Section \ref{sec:simulation} the comparison between the theory and direct numerical simulation of the underlying equations in Schrodinger representation has been presented. Finally, in Section \ref{sec:remarks} we draw some conclusions.

\section{\label{sec:sketch}Sketch of Derivation of the Quantum Single-Wave Model}

We start with the von Neumann-Maxwell equations in one spatial dimension
\begin{equation}
\partial_{t_1} F + {\frac {P_1} {m}} \partial_{X_1} F \nonumber
\end{equation}
\begin{equation}
= {\frac {i e} {\hbar}} {\left[ \phi {\left( X_1 + {\frac {i \hbar} {2}} \partial_{P_1} \right)} - \phi {\left( X_1 - {\frac {i \hbar} {2}} \partial_{P_1} \right)} \right]} {\left( F_0 + F \right)}, \label{VonNeumann}
\end{equation}
\begin{equation}
\partial_{X_1}^2 \phi = {\frac {e N_0} {\epsilon_0}} \int {\rm d} P_1 F {\left( X_1, P_1; t_1 \right)}. \label{Poisson}
\end{equation}
Here, $F_0 {\left( P_1 \right)} + F {\left( X_1, P_1; t_1 \right)}$ is the Wigner quasi-probability distribution function \cite{Wigner}, $m$ and $e$ are the electron rest mass and charge, respectively, and $\epsilon_0$ is the vacuum permittivity. Furthermore, $N_0$ is the density of the background ions, while all independent variables with index "$1$" refer to the real time and phase-space variables. The latter is adopted only for the sake of clarity in distinguishing the final notation for the independent variables in a sequence of scaling transformations performed throughout the derivation. Note that the quasi-probability distribution function $F {\left( X_1, P_1; t_1 \right)}$ is the departure from the equilibrium one $F_0 {\left( P_1 \right)}$. It is convenient to nondimensionalize the variables in the above Eqs. (\ref{VonNeumann}) and (\ref{Poisson}) according to
\begin{equation}
T = \omega_p t_1, \quad X = {\frac {\omega_p X_1} {v_c}}, \quad \Phi = {\frac {e \phi} {m v_c^2}}, \quad P = {\frac {P_1} {m v_c}}, \label{Nondimension}
\end{equation}
where $v_c$ is a velocity characteristic for the system, which is to a large extent arbitrary, while $\omega_p$ is the plasma frequency
\begin{equation}
\omega_p = {\frac {e^2 N_0} {m \epsilon_0}}. \label{PlasmaFreq}
\end{equation}
The von Neumann-Maxwell system of equations is cast now in the form
\begin{equation}
\partial_T F + P \partial_X F \nonumber
\end{equation}
\begin{equation}
= {\frac {i} {\hbar_1}} {\left[ \Phi {\left( X + {\frac {i \hbar_1} {2}} \partial_P \right)} - \Phi {\left( X - {\frac {i \hbar_1} {2}} \partial_P \right)} \right]} {\left( F_0 + F \right)}, \label{VonNeumann1}
\end{equation}
\begin{equation}
\partial_{X}^2 \Phi = \int {\rm d} P F {\left( X, P; T \right)}, \label{Poisson1}
\end{equation}
where the new dimensionless Planck's constant is given by the expression
\begin{equation}
\hbar_1 = {\frac {\hbar \omega_p} {m v_c^2}}. \label{Planck1}
\end{equation}
Using the formal Fourier representation of the Wigner quasi-probability distribution in momentum space, and performing simple algebraic manipulations, it is possible to cast the von Neumann equation in an equivalent form
\begin{equation}
\partial_T F + P \partial_X F \nonumber
\end{equation}
\begin{equation}
= {\frac {i} {2 \pi \hbar_1}} \int {\rm d} \lambda {\rm d} P_1 e^{i \lambda {\left( P - P_1 \right)}} \times \nonumber
\end{equation}
\begin{equation}
{\left[ \Phi {\left( X - {\frac {\hbar_1 \lambda} {2}} \right)} - \Phi {\left( X + {\frac {\hbar_1 \lambda} {2}} \right)} \right]} {\left[ F_0 {\left( P_1 \right)} + F {\left( X, P_1; T \right)} \right]}, \label{VonNeumannEquiv}
\end{equation}
which sometimes proves to be more convenient for direct analysis.

The linear stability of the equilibrium electron distribution $F_0 {\left( P \right)}$ is now determined by the roots of the dispersion function
\begin{equation}
{\cal D} {\left( k, c \right)} = \nonumber
\end{equation}
\begin{equation}
1 - {\frac {1} {\hbar_1 k^3}} \int \limits_{- \infty}^{\infty} {\frac {{\rm d} P} {P - c}} {\left[ F_0 {\left( P + {\frac {\hbar_1 k} {2}} \right)} - F_0 {\left( P - {\frac {\hbar_1 k} {2}} \right)} \right]}, \label{DispFunction}
\end{equation}
where $k$ is the wave number, and $c = \omega / k$. The quantum correction to the dispersion function (\ref{DispFunction}) as compared to the classical limiting case, when $\hbar_1 \rightarrow 0$, is of second order in $\hbar_1$
\begin{equation}
{\cal D} {\left( k, c \right)} = {\cal D}_{cl} {\left( k, c \right)} - {\frac {\hbar_1^2} {24}} \int \limits_{- \infty}^{\infty} {\frac {{\rm d} P} {P - c}} {\frac {{\rm d}^3 F_0 {\left( P \right)} } {{\rm d} P^3}} + \dots, \label{QuantumCorr}
\end{equation}
where
\begin{equation}
{\cal D}_{cl} {\left( k, c \right)} = 1 - {\frac {1} {k^2}} \int \limits_{- \infty}^{\infty} {\frac {{\rm d} P} {P - c}} {\frac {{\rm d} F_0 {\left( P \right)} } {{\rm d} P}}, \label{ClassicalDisp}
\end{equation}
is the classical dispersion function. Therefore, for sufficiently small $\hbar_1$ one should not expect significant modifications in the linear stability properties as described by the roots of the classical dispersion function (\ref{ClassicalDisp}).

Following del-Castillo-Negrete \cite{DelCastillo}, we assume the existence of a marginally stable state, prescribed by a stationary inflection point of the equilibrium distribution $F_0 {\left( P \right)}$. Since the analysis does not add anything new apart from taking into account the "quantum" Vlasov equation (\ref{VonNeumann1}) [or its equivalent form (\ref{VonNeumannEquiv})] instead of the classical one, we shall omit details of the calculation here. The final result is
\begin{equation}
\partial_t f + p \partial_x f = {\frac {i} {{\widetilde{\hbar}}}} {\left[ \varphi {\left( x + {\frac {i {\widetilde{\hbar}}} {2}} \partial_p \right)} - \varphi {\left( x - {\frac {i {\widetilde{\hbar}}} {2}} \partial_p \right)} \right]} f, \label{VonNeumannFin}
\end{equation}
\begin{equation}
\varphi {\left( x; t \right)} = a (t) e^{i x} + a^{\ast} (t) e^{- i x}, \label{Potential2}
\end{equation}
\begin{equation}
\sigma {\frac {{\rm d} a} {{\rm d} t}} + i l a = i {\left \langle f e^{- i x} \right \rangle}, \label{PotPoisson2}
\end{equation}
where the operator averaging is denoted by
\begin{equation}
{\left \langle \dots \right \rangle} = {\frac {1} {2 \pi}} \int \limits_{- \infty}^{\infty} {\rm d} p \int \limits_{0}^{2 \pi} {\rm d} x \dots. \label{Average}
\end{equation}
The dimensionless time $t$, coordinate $x$ and momentum $p$, as well as the single wave amplitude $a$ and the quasi-probability distribution function $f$ are appropriately nondimensionalized \cite{DelCastillo}. In addition, the new dimensionless Planck's constant ${\widetilde{\hbar}}$ is defined as
\begin{equation}
{\widetilde{\hbar}} = {\frac {2 \pi \hbar |\gamma|} {m v_c L |\lambda|}}, \label{Newhbar}
\end{equation}
where $L$ is the spatial length of the system, and
\begin{equation}
\gamma = k_{\ast}^2 \partial_c {\cal D} {\left( k_{\ast}, c_{\ast} \right)}, \qquad \quad \lambda = 2 \Lambda k_{\ast}^2. \label{AdditionalConst}
\end{equation}
In the above definition (\ref{AdditionalConst}), ${\left( k_{\ast}, c_{\ast} \right)}$ characterizes the marginally stable state in the ${\left( k, c \right)}$-space, and $\Lambda$ is a parameter measuring the difference between the domain length $L$ and the wavelength $2 \pi / k_{\ast}$ of the inflection point mode. To simplify notations in what follows, we shall drop the tilde over the ${\widetilde{\hbar}}$ as defined by Eq. (\ref{Newhbar}).

The linear stability analysis of the quantum single-wave model as described by the von-Neumann equation (\ref{VonNeumannFin}), coupled with the equations for the single field mode (\ref{Potential2}) and (\ref{PotPoisson2}) is presented in Appendix \ref{sec:appendix}.

\section{\label{sec:schrodpic}The Schrodinger Picture of the Quantum Single-Wave Model}

The von Neumann equation (also known as Liouville-von Neumann equation) describes how the Wigner quasi-probability distribution evolves in time, while the Schrodinger equation describes how pure states with support in configuration space evolve in time. In fact, the two equations are equivalent, in the sense that either can be derived from the other \cite{Schwabl}. More convenient for the subsequent analysis is the Schrodinger form of the system of equations (\ref{VonNeumannFin}) - (\ref{PotPoisson2}), which can be written as
\begin{equation}
i \hbar \partial_t \Psi = - {\frac {\hbar^2} {2}} \partial_x^2 \Psi - \varphi \Psi, \label{Schrodinger}
\end{equation}
\begin{equation}
\varphi {\left( x; t \right)} = a (t) e^{i x} + a^{\ast} (t) e^{- i x}, \label{Potential}
\end{equation}
\begin{equation}
\sigma {\frac {{\rm d} a} {{\rm d} t}} + i l a = i {\left \langle {\left| \Psi \right|}^2 e^{- i x} \right \rangle}, \label{PotPoisson}
\end{equation}
where now the averaging specified by Eq. (\ref{Average}) involves integration on the spatial variable $x$ only. It is worthwhile mentioning that an alternative quantum hydrodynamics approach (QHD) is possible \cite{Tsintsadze,Haas}, however for the present purposes we prefer to confine ourselves to the Schrodinger picture.

Taking into account the intrinsic periodicity of the system in the spatial variable $x$, we represent the wave function in the form
\begin{equation}
\Psi {\left( x; t \right)} = \sum \limits_{m = - \infty}^{\infty} \Psi_m {\left( t \right)} e^{i m x}. \label{Periodic}
\end{equation}
The dynamics of the Fourier harmonics $\Psi_m$ is governed by the following equation
\begin{equation}
{\left( i \hbar \partial_t - {\frac {\hbar^2 m^2} {2}} \right)} \Psi_m =  - a \Psi_{m-1} - a^{\ast} \Psi_{m+1}, \label{MFourierDyn}
\end{equation}
which should be supplemented by the equation
\begin{equation}
{\left( \sigma \partial_t + i l \right)} a =  i \sum \limits_{n = - \infty}^{\infty} \Psi_n \Psi_{n-1}^{\ast}, \label{SelfField}
\end{equation}
describing the evolution of the amplitude of the self-consistent electric field single-wave mode.

Let $k_0$ be a given harmonic number. Since direct interaction between waves as described by Eq. (\ref{SelfField}) involves only closest neighbors, it is straightforward to verify that an exact solution of the system of equations (\ref{MFourierDyn}) and (\ref{SelfField}) is of the form
\begin{equation}
\Psi_n^{(0)} = {\cal A}_n e^{- i \hbar n^2 t / 2}, \qquad n = \dots k_0 - 2, k_0, k_0 + 2, \dots, \label{ExactSol}
\end{equation}
and $\Psi_n^{(0)} = 0$ otherwise. In addition, the amplitude of the single wave mode vanishes ${\left( a^{(0)} = 0 \right)}$ as should be expected. This solution represents a superposition of non interacting plane waves, which remain uncoupled due to the fact that the nonzero wave harmonics are of either even or odd wave number, respectively.

To begin with, we consider the case, where the only wave with nonzero amplitude is the one with a prescribed harmonic number $s$ ${\left( {\cal A}_s \neq 0 \right)}$, while all others vanish. The solution we shall be seeking can be expressed as follows
\begin{equation}
\Psi_m = \Psi_m^{(0)} + \epsilon \Lambda_m, \qquad \quad a = \epsilon \alpha, \label{SolSought}
\end{equation}
where
\begin{equation}
\Psi_m^{(0)} = {\frac {\delta_{ms}} {\sqrt{2 \pi}}} e^{- i \hbar m^2 t / 2 + i \varphi_m}, \label{ZeroOrderPsim}
\end{equation}
and $\epsilon$ is a formal small parameter. Thus, we obtain
\begin{equation}
{\left( i \hbar \partial_t - {\frac {\hbar^2 m^2} {2}} \right)} \Lambda_m = - \epsilon {\left( \alpha \Lambda_{m-1} + \alpha^{\ast} \Lambda_{m+1} \right)}, \label{PertDynM}
\end{equation}
for $m \neq s-1$ or $m \neq s+1$, and
\begin{equation}
{\left[ i \hbar \partial_t + {\frac {\hbar^2} {2}} (s-1)^2 \right]} \Lambda_{s-1}^{\ast} = \alpha \Psi_s^{(0)\ast} + \epsilon {\left( \alpha^{\ast} \Lambda_{s-2}^{\ast} + \alpha \Lambda_s^{\ast} \right)}, \label{PertDynMSm1}
\end{equation}
\begin{equation}
{\left[ i \hbar \partial_t - {\frac {\hbar^2} {2}} (s+1)^2 \right]} \Lambda_{s+1} = - \alpha \Psi_s^{(0)} - \epsilon {\left( \alpha \Lambda_s + \alpha^{\ast} \Lambda_{s+2} \right)}, \label{PertDynMSp1}
\end{equation}
\begin{equation}
{\left( \sigma \partial_t + i l \right)} \alpha = i {\left( \Psi_s^{(0)} \Lambda_{s-1}^{\ast} + \Psi_s^{(0)\ast} \Lambda_{s+1} \right)} + i \epsilon \sum_n \Lambda_n \Lambda_{n-1}^{\ast}, \label{SelfFieldPert}
\end{equation}

It is convenient to introduce new variables defined according to the expressions
\begin{equation}
\xi = \Psi_s^{(0)} \Lambda_{s-1}^{\ast} + \Psi_s^{(0)\ast} \Lambda_{s+1}, \quad \zeta = \Psi_s^{(0)} \Lambda_{s-1}^{\ast} - \Psi_s^{(0)\ast} \Lambda_{s+1}, \label{NewVarXi}
\end{equation}
where
\begin{equation}
\Lambda_{s-1} = \pi \Psi_s^{(0)}{\left( \xi^{\ast} + \zeta^{\ast} \right)}, \qquad \Lambda_{s+1} = \pi \Psi_s^{(0)}{\left( \xi - \zeta \right)}, \label{NewVarXiZeta}
\end{equation}
Simple algebraic manipulations of Eq. (\ref{PertDynMSm1}) - (\ref{SelfFieldPert}) yield the following equations for the new variables $\xi$ and $\zeta$
\begin{equation}
{\left( i \hbar \partial_t - \hbar^2 s \right)} \xi + {\frac {\hbar^2} {2}} \zeta = \epsilon W_{\xi}, \label{EquationXi}
\end{equation}
\begin{equation}
{\left( i \hbar \partial_t - \hbar^2 s \right)} \zeta + {\frac {\hbar^2} {2}} \xi = {\frac {\alpha} {\pi}} + \epsilon W_{\zeta}, \label{EquationZeta}
\end{equation}
where
\begin{equation}
W_{\xi} = \alpha^{\ast} \Psi_s^{(0)} \Lambda_{s-2}^{\ast} + \alpha \Psi_s^{(0)} \Lambda_s^{\ast} - \alpha \Psi_s^{(0)\ast} \Lambda_s - \alpha^{\ast} \Psi_s^{(0)\ast} \Lambda_{s+2}, \label{WXi}
\end{equation}
\begin{equation}
W_{\zeta} = \alpha^{\ast} \Psi_s^{(0)} \Lambda_{s-2}^{\ast} + \alpha \Psi_s^{(0)} \Lambda_s^{\ast} + \alpha \Psi_s^{(0)\ast} \Lambda_s + \alpha^{\ast} \Psi_s^{(0)\ast} \Lambda_{s+2}. \label{WZeta}
\end{equation}
Eliminating the auxiliary variable $\zeta$, we obtain a single equation for $\xi$
\begin{equation}
{\left[ {\left( \sigma \partial_t + i l \right)} {\left( i \partial_t - \hbar s \right)}^2 - {\frac {\hbar^2} {4}} {\left( \sigma \partial_t + i l \right)} + {\frac {i} {2 \pi}} \right]} \xi \nonumber
\end{equation}
\begin{equation}
= - {\frac {i \epsilon} {2 \pi}} \sum_n \Lambda_n \Lambda_{n-1}^{\ast} + {\frac {\epsilon} {\hbar}} {\left( \sigma \partial_t + i l \right)} {\left[ {\left( i \partial_t - \hbar s \right)} W_{\xi} - {\frac {\hbar} {2}} W_{\zeta} \right]}, \label{BasicEqXi}
\end{equation}
which should be supplemented by Eq. (\ref{SelfFieldPert}), written as
\begin{equation}
{\left( \sigma \partial_t + i l \right)} \alpha = i \xi + i \epsilon \sum_n \Lambda_n \Lambda_{n-1}^{\ast}. \label{SelfFieldBasic}
\end{equation}

\section{\label{sec:perturbative}Perturbative Solution of the Evolution Equations}

Following the standard perturbation approximation procedure \cite{TzenovBOOK}, we represent all dynamical variables as a series expansion in the formal small parameter $\epsilon$. In order to eliminate resonant (secular) terms appearing in successive orders, we define different time scales as prescribed by the method of multiple scales \cite{TzenovBOOK}. To zero order, the perturbation equations (\ref{BasicEqXi}) and (\ref{SelfFieldBasic}) can be written as
\begin{equation}
{\left[ {\left( \sigma \partial_t + i l \right)} {\left( i \partial_t - \hbar s \right)}^2 - {\frac {\hbar^2} {4}} {\left( \sigma \partial_t + i l \right)} + {\frac {i} {2 \pi}} \right]} \xi_0 = 0, \label{ZeroBasicEqXi}
\end{equation}
\begin{equation}
{\left( \sigma \partial_t + i l \right)} \alpha_0 = i \xi_0. \label{ZeroSelfFieldBasic}
\end{equation}
The general solution of Eq. (\ref{ZeroBasicEqXi}) can be expressed as
\begin{equation}
\xi_0 {\left( t, t_1, t_2, \dots \right)} = \sum \limits_{\beta = 1}^{3} {\cal B}_{\beta} {\left( t_1, t_2, \dots \right)} e^{i \omega_{\beta} t}, \label{ZeroOrderSol}
\end{equation}
where $t_1 = \epsilon t$, $t_2 = \epsilon^2 t$, and so on are slow times, and $\omega_{\beta}$ are the three roots of the characteristic equation
\begin{equation}
{\left( \sigma \omega + l \right)} {\left( \omega + \hbar s \right)}^2 - {\frac {\hbar^2} {4}} {\left( \sigma \omega + l \right)} + {\frac {1} {2 \pi}} = 0. \label{CharacteristicEq}
\end{equation}
The zero-order amplitude of the single-wave mode $\alpha_0$ can be determined immediately from Eq. (\ref{ZeroSelfFieldBasic}) to be
\begin{equation}
\alpha_0 = \sum \limits_{\beta = 1}^{3} {\frac {{\cal B}_{\beta}} {\sigma \omega_{\beta} + l}} e^{i \omega_{\beta} t}. \label{ZeroOrderAlpha}
\end{equation}
From Eq. (\ref{EquationXi}) for the auxiliary variable $\zeta_0$, we obtain
\begin{equation}
\zeta_0 = {\frac {2} {\hbar}} \sum \limits_{\beta = 1}^{3} {\left( \omega_{\beta} + \hbar s \right)} {\cal B}_{\beta} e^{i \omega_{\beta} t}, \label{ZeroOrderZeta}
\end{equation}
For the perturbed harmonics of the wave function we can now write
\begin{equation}
\Lambda_{s-1}^{(0)} = {\sqrt{\frac {\pi} {2}}} \sum \limits_{\beta = 1}^{3} \Lambda_{\beta}^{(-)} {\cal B}_{\beta}^{\ast} e^{-i {\left( \omega_{\beta}^{\ast} + \hbar s^2 / 2 \right)} t  + i \varphi_s}, \label{ZeroOrderSm1}
\end{equation}
\begin{equation}
\Lambda_{s+1}^{(0)} = {\sqrt{\frac {\pi} {2}}} \sum \limits_{\beta = 1}^{3} \Lambda_{\beta}^{(+)} {\cal B}_{\beta} e^{i {\left( \omega_{\beta} - \hbar s^2 / 2 \right)} t  + i \varphi_s}, \label{ZeroOrderSp1}
\end{equation}
where the coefficients in the above expressions are given by
\begin{equation}
\Lambda_{\beta}^{(-)} = 1 + {\frac {2} {\hbar}} {\left( \omega_{\beta}^{\ast} + \hbar s \right)}, \qquad \Lambda_{\beta}^{(+)} = 1 - {\frac {2} {\hbar}} {\left( \omega_{\beta} + \hbar s \right)}. \label{LambdaBetaPM}
\end{equation}
Note that in zero order all other harmonics as described by Eq. (\ref{PertDynM}) vanish ${\left( \Lambda_m^{(0)} = 0, \quad m \neq s-1, s+1 \right)}$.

Even in the lowest order an interesting feature of quantum single-wave model can be observed. Two adjacent harmonics (with wave numbers $s-1$ and $s+1$) are generated on either side of the initial isolated harmonic with wave number $s$. This excitation is due to the closest neighbor interaction, characterized by the evolution law of the amplitude of the self-consistent electric potential. As it will become clear from the subsequent exposition, once harmonics $s-1$ and $s+1$ are generated, they give rise to their closest neighbors $s-2$ and $s+2$, respectively, and so on until all the spectrum becomes eventually populated. This phenomenon has also been detected in the numerical simulations.

Let us proceed with the first order in the formal expansion parameter $\epsilon$. Since the quantity $\xi_1$ satisfies equation similar to Eq. (\ref{ZeroBasicEqXi}), without loss of generality it can be assumed $\xi_1 = 0$. This immediately leads to
\begin{equation}
\zeta_1 = 0, \qquad \qquad \alpha_1 = 0. \label{FirstOrderSol}
\end{equation}
Thus, the first order corrections to the $(s-1)$-st and $(s+1)$-st harmonics of the wave function also vanish
\begin{equation}
\Lambda_{s-1}^{(1)} = 0, \qquad \qquad \Lambda_{s+1}^{(1)} = 0. \label{FirstOrderSmp1}
\end{equation}
Additional consequence of the vanishing first-order solution $\xi_1 = 0$ is the fact that the amplitudes ${\cal B}_{\beta}$ introduced in Eq. (\ref{ZeroOrderSol}) do not depend on the time scale $t_1$ ${\left( \partial_{t_1} \xi_0 = 0 \right)}$.

Taking into account Eq. (\ref{PertDynM}) and the fact that the arbitrary phase $\varphi_s$ may depend on slower time scales, we obtain
\begin{equation}
{\left( i \hbar \partial_t - {\frac {\hbar^2 s^2} {2}} \right)} \Lambda_s^{(1)} + i \hbar \partial_{t_1} \Psi_s^{(0)} = - \alpha_0 \Lambda_{s-1}^{(0)} - \alpha_0^{\ast} \Lambda_{s+1}^{(0)}. \label{FirstEqLambdaS}
\end{equation}
The right-hand-side of the above equation contains resonant (secular) terms, which can be eliminated by equating them to the second term on the left-hand-side. Thus, we obtain an equation describing the phase dynamics of the basic harmonic $\Psi_s^{(0)}$ [see Eq. (\ref{ZeroOrderPsim})]. In the case, where all three roots of the characteristic equation (\ref{CharacteristicEq}) are real, the amplitude dependent phase shift is governed by the following equation
\begin{equation}
\partial_t \varphi_s = {\frac {2 \pi} {\hbar}} \sum \limits_{\beta = 1}^{3} {\frac {{\left| {\cal B}_{\beta} \right|}^2} {\sigma \omega_{\beta} + l}}. \label{PhaseShiftR}
\end{equation}
If $\omega_1$ is a real root, and $\omega_2 = \omega_3^{\ast}$ is a complex conjugate pair, the equation for the phase shift can be written as follows
\begin{equation}
\partial_t \varphi_s = {\frac {2 \pi} {\hbar}} {\left( {\frac {{\left| {\cal B}_1 \right|}^2} {\sigma \omega_1 + l}} + {\frac {{\cal B}_2 {\cal B}_3^{\ast}} {\sigma \omega_2 + l}} + c.c. \right)}. \label{PhaseShiftI}
\end{equation}
Remarkably enough, in both cases the slowly varying phase of the basic harmonic $s$ is a real number. Solving now Eq. (\ref{FirstEqLambdaS}) for the first-order correction to the amplitude of the basic harmonic, we obtain
\begin{equation}
\Lambda_s^{(1)} = {\frac {1} {\hbar}} {\sqrt{\frac {\pi} {2}}} {\sum_{\beta \gamma}}^{\prime} {\frac {{\cal B}_{\beta} {\cal B}_{\gamma}^{\ast}} {\omega_{\beta} - \omega_{\gamma}^{\ast}}} {\left( {\frac {\Lambda_{\gamma}^{(-)}} {\sigma \omega_{\beta} + l}} + {\frac {\Lambda_{\beta}^{(+)}} {\sigma \omega_{\gamma}^{\ast} + l}} \right)} \nonumber
\end{equation}
\begin{equation}
\times e^{i {\left( \omega_{\beta} - \omega_{\gamma}^{\ast} - \hbar s^2 / 2 \right)} t + i \varphi_s}. \label{FirstOrdSolLS}
\end{equation}
The prime over the summation symbol in Eq. (\ref{FirstOrdSolLS}) implies that terms for which $\omega_{\beta} = \omega_{\gamma}^{\ast}$ are excluded. Note that such terms have already been taken into account in writing the amplitude-phase equations (\ref{PhaseShiftR}) and (\ref{PhaseShiftI}), respectively.

As already mentioned above, a new harmonic pair with wave numbers $s-2$ and $s+2$ is generated in first order. From Eq. (\ref{PertDynM}) we obtain
\begin{equation}
{\left[ i \hbar \partial_t - {\frac {\hbar^2} {2}} {\left( s - 2 \right)}^2 \right]} \Lambda_{s-2}^{(1)} = - \alpha_0^{\ast} \Lambda_{s-1}^{(0)}, \label{FirstEqLambdaSm2}
\end{equation}
\begin{equation}
{\left[ i \hbar \partial_t - {\frac {\hbar^2} {2}} {\left( s + 2 \right)}^2 \right]} \Lambda_{s+2}^{(1)} = - \alpha_0 \Lambda_{s+1}^{(0)}. \label{FirstEqLambdaSp2}
\end{equation}
The solutions of the last two equations are regular, and can be expressed as
\begin{equation}
\Lambda_{s-2}^{(1)} = - {\frac {1} {\hbar}} {\sqrt{\frac {\pi} {8}}} \sum_{\beta \gamma} F_{\beta \gamma} {\cal B}_{\beta}^{\ast} {\cal B}_{\gamma}^{\ast} e^{-i {\left( \omega_{\beta}^{\ast} + \omega_{\gamma}^{\ast} + \hbar s^2 / 2 \right)} t + i \varphi_s}, \label{FirstOrdSolLSm2}
\end{equation}
\begin{equation}
\Lambda_{s+2}^{(1)} = {\frac {1} {\hbar}} {\sqrt{\frac {\pi} {8}}} \sum_{\beta \gamma} G_{\beta \gamma} {\cal B}_{\beta} {\cal B}_{\gamma} e^{i {\left( \omega_{\beta} + \omega_{\gamma} - \hbar s^2 / 2 \right)} t + i \varphi_s}, \label{FirstOrdSolLSp2}
\end{equation}
where the symmetric matrices $F_{\beta \gamma}$ and $G_{\beta \gamma}$ are given by the expressions
\begin{equation}
F_{\beta \gamma} = {\frac {1} {\omega_{\beta}^{\ast} + \omega_{\gamma}^{\ast} + 2 \hbar (s-1)}} {\left( {\frac {\Lambda_{\beta}^{(-)}} {\sigma \omega_{\gamma}^{\ast} + l}} + {\frac {\Lambda_{\gamma}^{(-)}} {\sigma \omega_{\beta}^{\ast} + l}} \right)}, \label{SymmetricF}
\end{equation}
\begin{equation}
G_{\beta \gamma} = {\frac {1} {\omega_{\beta} + \omega_{\gamma} + 2 \hbar (s+1)}} {\left( {\frac {\Lambda_{\beta}^{(+)}} {\sigma \omega_{\gamma} + l}} + {\frac {\Lambda_{\gamma}^{(+)}} {\sigma \omega_{\beta} + l}} \right)}. \label{SymmetricG}
\end{equation}

\section{\label{sec:amplitude}The Amplitude Equation}

The final step of our perturbation analysis consists in deriving an amplitude equation for the slowly varying envelope functions ${\cal B}_{\beta}$, introduced in the zero-order approximation (\ref{ZeroOrderSol}). For that purpose, let us write the second-order equation following from the expression (\ref{BasicEqXi})
\begin{equation}
{\left[ {\left( \sigma \partial_t + i l \right)} {\left( i \partial_t - \hbar s \right)}^2 - {\frac {\hbar^2} {4}} {\left( \sigma \partial_t + i l \right)} + {\frac {i} {2 \pi}} \right]} \xi_2 \nonumber
\end{equation}
\begin{equation}
+ {\left[ \sigma {\left( i \partial_t - \hbar s \right)}^2 + 2 i {\left( \sigma \partial_t + i l \right)} {\left( i \partial_t - \hbar s \right)} - {\frac {\sigma \hbar^2} {4}} \right]} \partial_{t_2} \xi_0 \nonumber
\end{equation}
\begin{equation}
= - {\frac {1} {2 \pi}} {\left( \Lambda_{s-1}^{(0)} \Lambda_{s-2}^{(1)\ast} + \Lambda_{s+1}^{(0)} \Lambda_s^{(1)\ast} + \Lambda_{s-1}^{(0)\ast} \Lambda_s^{(1)} + \Lambda_{s+1}^{(0)\ast} \Lambda_{s+2}^{(1)} \right)} \nonumber
\end{equation}
\begin{equation}
+ {\frac {1} {\hbar}} {\left( \sigma \partial_t + i l \right)} {\left[ {\left( i \partial_t - \hbar s \right)} W_{\xi}^{(1)} - {\frac {\hbar} {2}} W_{\zeta}^{(1)} \right]}. \label{SecondBasicEqXi}
\end{equation}
After substituting into the right-hand-side of the above equation the already found solutions in previous orders and performing simple but rather cumbersome algebraic manipulations, one can isolate the secular terms. These can be canceled by equating them to the second term on the left-hand-side of Eq. (\ref{SecondBasicEqXi}). The result represents the amplitude equation governing the evolution of the slowly varying amplitudes ${\cal B}_{\beta}$ on a time scale of the order $1 / \epsilon^2$-times longer as compared to the fast oscillations with frequencies specified by the characteristic equation (\ref{CharacteristicEq}).

In the case, where all three roots of the characteristic equation (\ref{CharacteristicEq}) are real, we obtain
\begin{equation}
i \hbar V_{\beta} \partial_t {\cal B}_{\beta} = - {\frac {1} {4}} \sum \limits_{\gamma = 1}^{3} \Gamma_{\beta \gamma} {\cal B}_{\beta} {\left| {\cal B}_{\gamma} \right|}^2 - {\frac {1} {4}} \sum \limits_{\gamma \neq \beta} {\frac {\Delta_{\beta \gamma}} {\omega_{\beta} - \omega_{\gamma}}} {\cal B}_{\beta} {\left| {\cal B}_{\gamma} \right|}^2, \label{AmplEqReal}
\end{equation}
where
\begin{equation}
V_{\beta} = \sigma {\left( \omega_{\beta} + \hbar s \right)}^2 + 2 {\left( \sigma \omega_{\beta} + l \right)} {\left( \omega_{\beta} + \hbar s \right)} - {\frac {\sigma \hbar^2} {4}}, \label{GroupVelocity}
\end{equation}
and
\begin{equation}
\Gamma_{\beta \gamma} = F_{\beta \gamma} \Lambda_{\gamma}^{(-)} - G_{\beta \gamma} \Lambda_{\gamma}^{(+)} \nonumber
\end{equation}
\begin{equation}
+ {\frac {\sigma \omega_{\beta} + l} {\hbar {\left( \sigma \omega_{\gamma} + l \right)}}} {\left[ 2 {\left( \omega_{\beta} + \hbar s \right)} {\left( F_{\beta \gamma} + G_{\beta \gamma} \right)} + \hbar {\left( F_{\beta \gamma} - G_{\beta \gamma} \right)} \right]}, \label{CoeffReal1}
\end{equation}
\begin{equation}
\Delta_{\beta \gamma} = {\frac {\Lambda_{\beta}^{(-)} \Lambda_{\gamma}^{(+)} - \Lambda_{\beta}^{(+)} \Lambda_{\gamma}^{(-)}} {\sigma \omega_{\gamma} + l}} + {\frac {\Lambda_{\gamma}^{(+)2} - \Lambda_{\gamma}^{(-)2}} {\sigma \omega_{\beta} + l}} \nonumber
\end{equation}
\begin{equation}
+ {\frac {4 {\left( \sigma \omega_{\beta} + l \right)}} {\hbar {\left( \sigma \omega_{\gamma} + l \right)}}} {\left( 2 {\frac {\omega_{\beta} + \hbar s} {\sigma \omega_{\gamma} + l}} + {\frac {\omega_{\beta} - \omega_{\gamma}} {\sigma \omega_{\beta} + l}}  \right)}. \label{CoeffReal2}
\end{equation}
Similar amplitude equations can be obtained when one of the roots of the characteristic equation is real, while the other two comprise a complex conjugate pair. In this case, however, the exponential growth of the Fourier harmonics determined from zero and first order is dominating and the nonlinear wave interaction yields relatively small contribution.

The terms proportional to $\Gamma_{\beta \beta}$ describe the effects of phase self-modulation, whereas those proportional to $\Gamma_{\beta \gamma}$ (with $\beta \neq \gamma$) are responsible for phase cross modulation. Both result in an intensity-dependent nonlinear frequency shift of the eigenfrequencies $\omega_{\beta}$, an effect similar to the nonlinear wave-number shift in nonlinear optics \cite{Agrawal}. Note also that the second term on the right-hand-side of Eq. (\ref{AmplEqReal}) describes purely cross-modulation effects. In addition, the four-wave mixing effect responsible for energy exchange between the three eigenmodes has been neglected. 

\section{\label{sec:standing}Nonlinear Standing Waves}

Let us assume a trivial time dependence of the wavefunction $\Psi$ of the form
\begin{equation}
\Psi {\left( x, t \right)} = \Phi {\left( x \right)} e^{- i \omega t}, \label{StandWaveForm}
\end{equation}
where $\Phi {\left( x \right)}$ is a real valued function of the position variable. According to Eq. (\ref{PotPoisson}) the single-wave amplitude $a$ has a stationary solution, which can be expressed as
\begin{equation}
a = {\frac {1} {l}} {\left \langle \Phi^2 {\left( x \right)} e^{- i x} \right \rangle}. \label{SingleWaveStation}
\end{equation}
Substituting the above expressions into the Schrodinger equation (\ref{Schrodinger}), we obtain
\begin{equation}
\partial_x^2 \Phi + \lambda \Phi + \mu {\left[ e^{i x} {\left \langle \Phi^2 {\left( \xi \right)} e^{- i \xi} \right \rangle} + e^{- i x} {\left \langle \Phi^2 {\left( \xi \right)} e^{i \xi} \right \rangle} \right]} \Phi = 0, \label{SchrodingerStation}
\end{equation}
where
\begin{equation}
\lambda = {\frac {2 \omega} {\hbar}}, \qquad \qquad \mu = {\frac {2} {l \hbar^2}}. \label{ParameterStation}
\end{equation}
This latter equation describes the standing wave solutions of the time-dependent equation (\ref{Schrodinger}), which are the states with definite energy, instead of a probability distribution of different energies as discussed in the previous three sections.

Taking into account the periodicity of the problem, it is natural to represent the solution of Eq. (\ref{SchrodingerStation}) as a Fourier series
\begin{equation}
\Phi {\left( x \right)} = \sum \limits_{m = - \infty}^{\infty} \Phi_m e^{imx}, \label{FourierStation}
\end{equation}
similar to the representation given by Eq. (\ref{Periodic}). Thus for the Fourier amplitudes, we obtain
\begin{equation}
{\left( m^2 - \lambda \right)} \Phi_m \nonumber
\end{equation}
\begin{equation}
= \mu {\left( \Phi_{m-1} \sum_n \Phi_n \Phi_{n-1}^{\ast} + \Phi_{m+1} \sum_n \Phi_n^{\ast} \Phi_{n-1} \right)}. \label{FourierAmplEqn}
\end{equation}
Although it is in principle possible to solve the nonlinear equations (\ref{FourierAmplEqn}) for a finite number of Fourier amplitudes, it is instructive to consider the perturbative solution discussed below.

The zero-order solution is obvious
\begin{equation}
\Phi_m^{(0)} = A \delta_{ms}, \qquad \qquad \lambda_0 = s^2, \label{ZeroSolStat}
\end{equation}
where $s$ is a particular Fourier harmonic number and $A$ is an arbitrary complex constant. In first order, we obtain
\begin{equation}
{\left( m^2 - s^2 \right)} \Phi_m^{(1)} - \lambda_1 \Phi_m^{(0)} \nonumber
\end{equation}
\begin{equation}
= \mu \Phi_{m-1}^{(0)} {\left( A \Phi_{s-1}^{(1)\ast} + A^{\ast} \Phi_{s+1}^{(1)} \right)} \nonumber
\end{equation}
\begin{equation}
+ \mu \Phi_{m+1}^{(0)} {\left( A^{\ast} \Phi_{s-1}^{(1)} + A \Phi_{s+1}^{(1)\ast} \right)}. \label{FirstEqnStat}
\end{equation}
From the equation for $\Phi_s^{(1)}$ it follows that $\lambda_1 = 0$ and $\Phi_s^{(1)} = A_1$, where $A_1$ is a new constant, which can eventually be set to zero. As before (see previous sections for comparison), the only nonzero Fourier harmonics appearing in first order are the ${\left( s \pm 1 \right)}$-st ones. These are linearly coupled via the equations
\begin{equation}
{\left( 2s - 1 + \mu {\left| A \right|}^2 \right)} \Phi_{s-1}^{(1)} + \mu A^2 \Phi_{s+1}^{(1)\ast} = 0, \label{FirstLinCoupStat1}
\end{equation}
\begin{equation}
- \mu A^{\ast 2} \Phi_{s-1}^{(1)} + {\left( 2s + 1 - \mu {\left| A \right|}^2 \right)} \Phi_{s+1}^{(1)\ast} = 0. \label{FirstLinCoupStat2}
\end{equation}
The above system of linear homogeneous equations has a non trivial solution, provided its determinant is zero. This gives
\begin{equation}
{\left| A \right|}^2 = {\frac {1 - 4 s^2} {2 \mu}}. \label{FirstAStat}
\end{equation}
Clearly, the parameter $\mu$ must be negative, which is equivalent to the condition $l = -1$. For the ${\left( s \pm 1 \right)}$-st Fourier amplitudes, we obtain
\begin{equation}
\Phi_{s+1}^{(1)} = B, \qquad \quad \Phi_{s-1}^{(1)} = {\frac {2 \mu A^2 B^{\ast}} {(2s - 1)^2}}, \label{FirstBStat}
\end{equation}
where again $B$ is an arbitrary complex constant. The second-order perturbation equation can be expressed as
\begin{equation}
{\left( m^2 - s^2 \right)} \Phi_m^{(2)} - \lambda_2 \Phi_m^{(0)} \nonumber
\end{equation}
\begin{equation}
= \mu \Phi_{m-1}^{(0)} {\left( A \Phi_{s-1}^{(2)\ast} + A^{\ast} \Phi_{s+1}^{(2)} \right)} \nonumber
\end{equation}
\begin{equation}
+ \mu \Phi_{m-1}^{(1)} {\left( A \Phi_{s-1}^{(1)\ast} + A^{\ast} \Phi_{s+1}^{(1)} \right)} \nonumber
\end{equation}
\begin{equation}
+ \mu \Phi_{m+1}^{(0)} {\left( A^{\ast} \Phi_{s-1}^{(2)} + A \Phi_{s+1}^{(2)\ast} \right)} \nonumber
\end{equation}
\begin{equation}
+ \mu \Phi_{m+1}^{(1)} {\left( A^{\ast} \Phi_{s-1}^{(1)} + A \Phi_{s+1}^{(1)\ast} \right)}. \label{SecondEqnStat}
\end{equation}
The nonlinear frequency shift $\lambda_2$ and the additional ${\left( s \pm 2 \right)}$-nd Fourier amplitudes appearing in second order can be found to be
\begin{equation}
\lambda_2 = - {\frac {4 \mu {\left| B \right|}^2} {(2 s - 1)^2}}, \label{NonlinFreqStat}
\end{equation}
\begin{equation}
\Phi_{s+2}^{(2)} = - {\frac {\mu A^{\ast} B^2} {2 (2 s - 1) (s+1)}}, \quad \Phi_{s-2}^{(2)} = {\frac {\mu^2 A^3 B^{\ast 2}} {(2 s - 1)^3 (s-1)}}. \label{SecondSolStat}
\end{equation}

Note that the arbitrary to this end complex constants $A_1$ and $B$ can be used to satisfy the normalization condition of the wave function. It is also important to mention at this point the following. The system (\ref{FourierAmplEqn}) comprises a set of $n$ nonlinear equations for a finite number $n$ of Fourier harmonics with $\lambda$ being kept as a free parameter. The latter can be determined from the normalization condition of the wave function. This usually leads to a nonlinear dispersion relation of $n$-th order, the roots of which determine the stability properties of the standing wave solution.

\section{\label{sec:simulation}Numerical Simulations}

In this section some of the predictions of the perturbation theory are benchmarked to the exact system dynamics obtained by solving Eqs. (\ref{Schrodinger}) - (\ref{PotPoisson}) numerically by means of the standard split-step Fourier method \cite{Agrawal}. Depending on the values of the relevant parameters the characteristic equation (\ref{CharacteristicEq}) can either posses three real roots or one real and two complex-conjugate roots.

\begin{figure}
\begin{center}
\includegraphics[width=8.0cm]{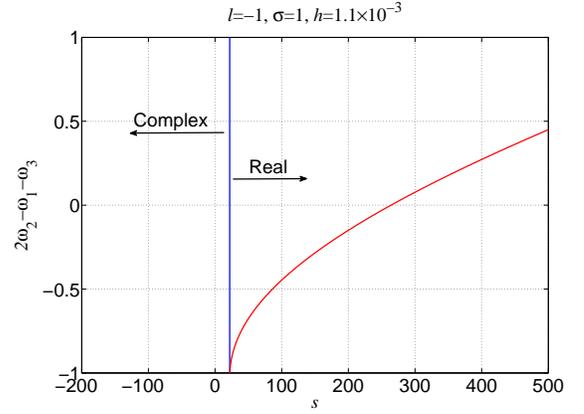}
\caption{\label{fig:epsart1} Regions of complex $s < 22$ and real $s > 22$ roots of the characteristic equation (\ref{CharacteristicEq}). The four-wave mixing (FWM) phase mismatch parameter $\delta \omega = 2 \omega_2 - \omega_1 - \omega_3$, where $\omega_1 < \omega_2 < \omega_3$ is also shown. }
\end{center}
\end{figure}

This is illustrated in Fig. \ref{fig:epsart1}, where the regions of three real and one real and two complex conjugate roots are marked for a particular choice of parameter values ${\left( \sigma, \quad l \quad {\rm and} \quad \hbar \right)}$. In addition, in the region of real roots the value of the phase mismatch parameter  $\delta \omega = 2 \omega_2 - \omega_1 - \omega_3$ responsible for the four-wave mixing (FWM) process \cite{Agrawal} is also shown. Within the limits of validity of the assumption for weak nonlinearity no evidence of energy exchange between the three frequency components of either $\Lambda_{s-1}^{(0)}$  or $\Lambda_{s+1}^{(0)}$  was found, which justifies neglecting the FWM in the derivation of Eq. (\ref{AmplEqReal}). This may be explained with the possible low value of the coefficient in front of the FWM term in the slowly-varying envelope approximation (SVEA) equation (\ref{AmplEqReal}). A similar situation exists in the classical case of the single-wave model \cite{MarTze}.

\begin{figure}
\begin{center}
\includegraphics[width=8.0cm]{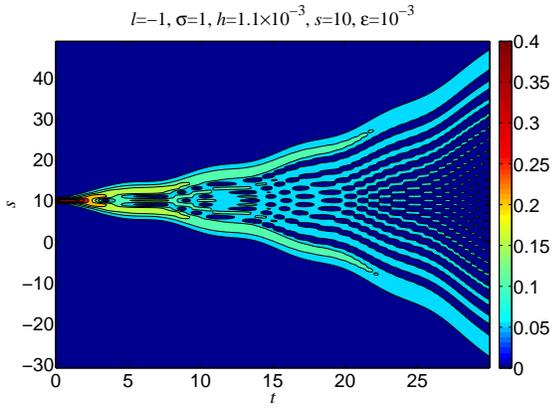}
\caption{\label{fig:epsart2} Evolution of the modulus of the spectrum of the wave function ${\left| \Psi(s,t) \right|}$ with time. Initial conditions in the unstable region of parameter space have been chosen and interpolation has been used along the s-axis.}
\end{center}
\end{figure}

\begin{figure}
\begin{center}
\includegraphics[width=8.0cm]{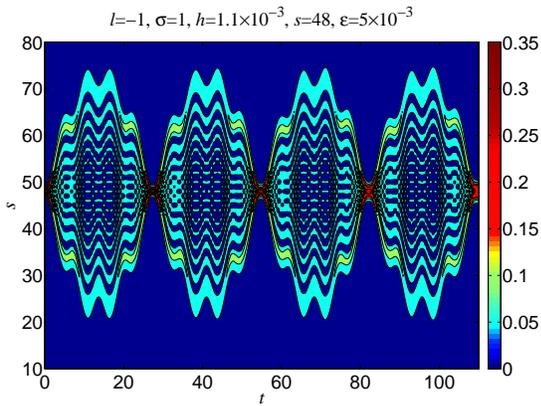}
\caption{\label{fig:epsart3} The same as in Fig. \ref{fig:epsart2} but the input is a stable modal triplet of the form (\ref{Triplet}).}
\end{center}
\end{figure}

As equations (\ref{ZeroOrderSm1}) and (\ref{ZeroOrderSp1}) show the amplitudes of the two weak sideband components  $\Lambda_{s-1}^{(0)}$ and $\Lambda_{s+1}^{(0)}$  either oscillate between their minimum and maximum values (three real roots), or alternatively, increase exponentially due to energy transfer from the powerful "pump" component $\Psi_{s}^{(0)}$  (two complex-conjugate roots of the characteristic equation). Thus modes excited in the region of purely real roots can be regarded as stable, whereas modes in the region of two complex-conjugate roots - as unstable. This observation is illustrated in Figures \ref{fig:epsart2} and \ref{fig:epsart3}.

Equations (\ref{Schrodinger}) - (\ref{PotPoisson}) have been solved with initial conditions in a triplet form
\begin{equation}
\Psi {\left( x, t=0 \right)} = {\frac {1} {\sqrt{2 \pi}}} {\left[ e^{isx} + \epsilon {\left( e^{i (s-1) x} + e^{i (s+1) x} \right)} \right]}. \label{Triplet}
\end{equation}
As Fig. \ref{fig:epsart2} shows exciting an unstable modal triplet (with $s<22$) leads to a gradual energy transfer to practically all neighboring modes and the excitation soon occupies the entire spectral range simulated numerically. In contrast, in the stable region energy is initially transferred to neighboring modes but then returns back to the mode that has been excited originally and the process repeats itself periodically in time.

\begin{figure}
\begin{center}
\includegraphics[width=8.0cm]{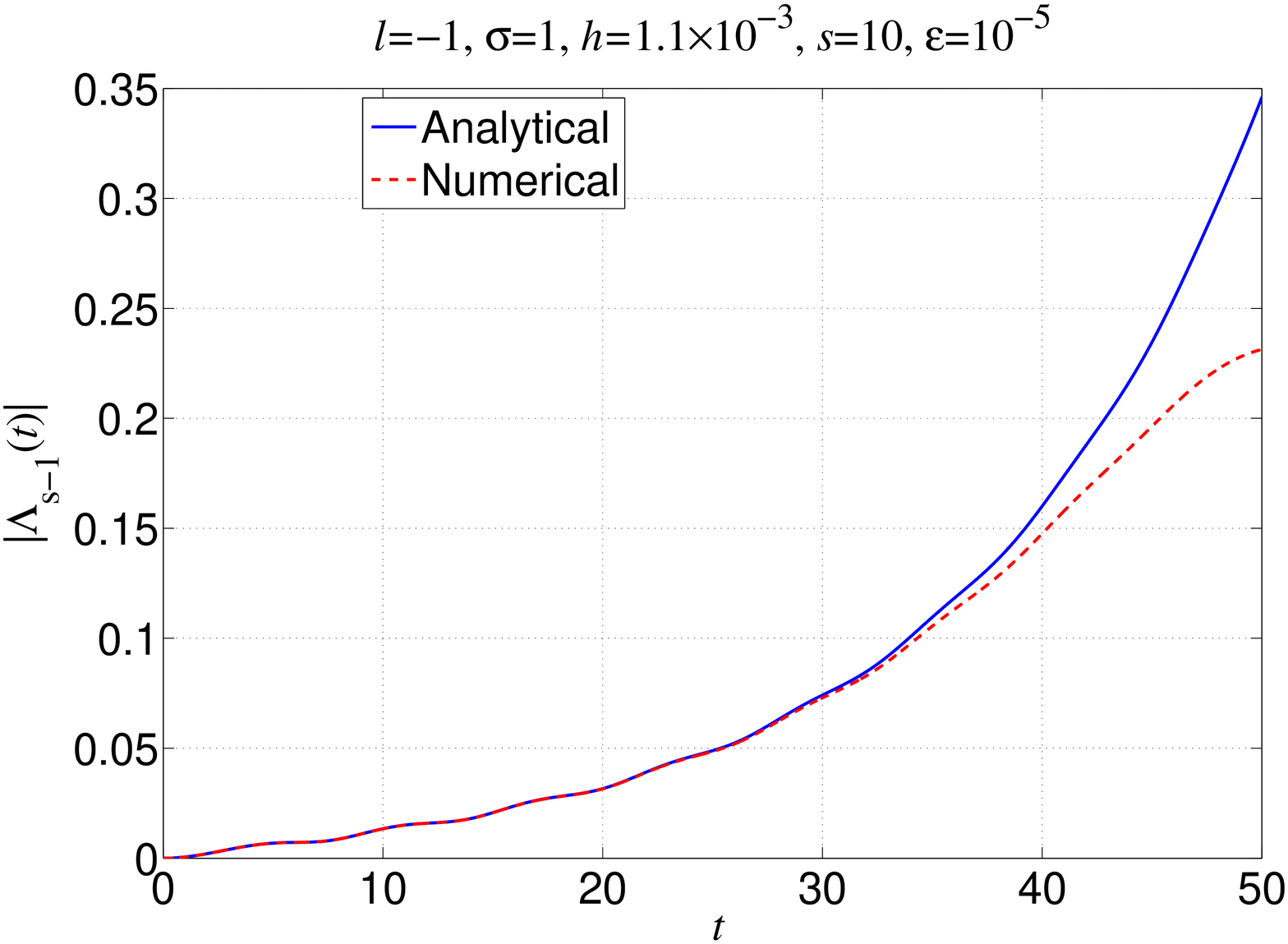}
\caption{\label{fig:epsart4} Comparison between the time-domain numerical and analytical [Eq. (\ref{ZeroOrderSm1})] solutions for $\Lambda_{s-1}^{(0)}$ in the unstable region.}
\end{center}
\end{figure}

In Figure \ref{fig:epsart4} the analytical expression for the sideband component amplitude ${\left| \Lambda_{s-1}^{(0)} \right|}$ obtained by means of a perturbative approach is compared to the exact numerical result. The choice of parameter values corresponds to the unstable mode region. As can be seen excellent agreement between the two results exists at earlier times, whereas at later times they deviate from each other. This is because the perturbative solution (\ref{ZeroOrderSm1}) is valid provided that ${\left| \Lambda_{s-1}^{(0)} \right|} \ll {\left| \Psi_{s}^{(0)} \right|}$ (weak probe interacting with a strong pump), which is indeed the case initially. However, since ${\left| \Lambda_{s-1}^{(0)} \right|}$ increases exponentially, the weak probe assumption is violated for sufficiently late times, which explains the observed difference between the analytical and the numerical result at later times.

\begin{figure}
\begin{center}
\includegraphics[width=8.0cm]{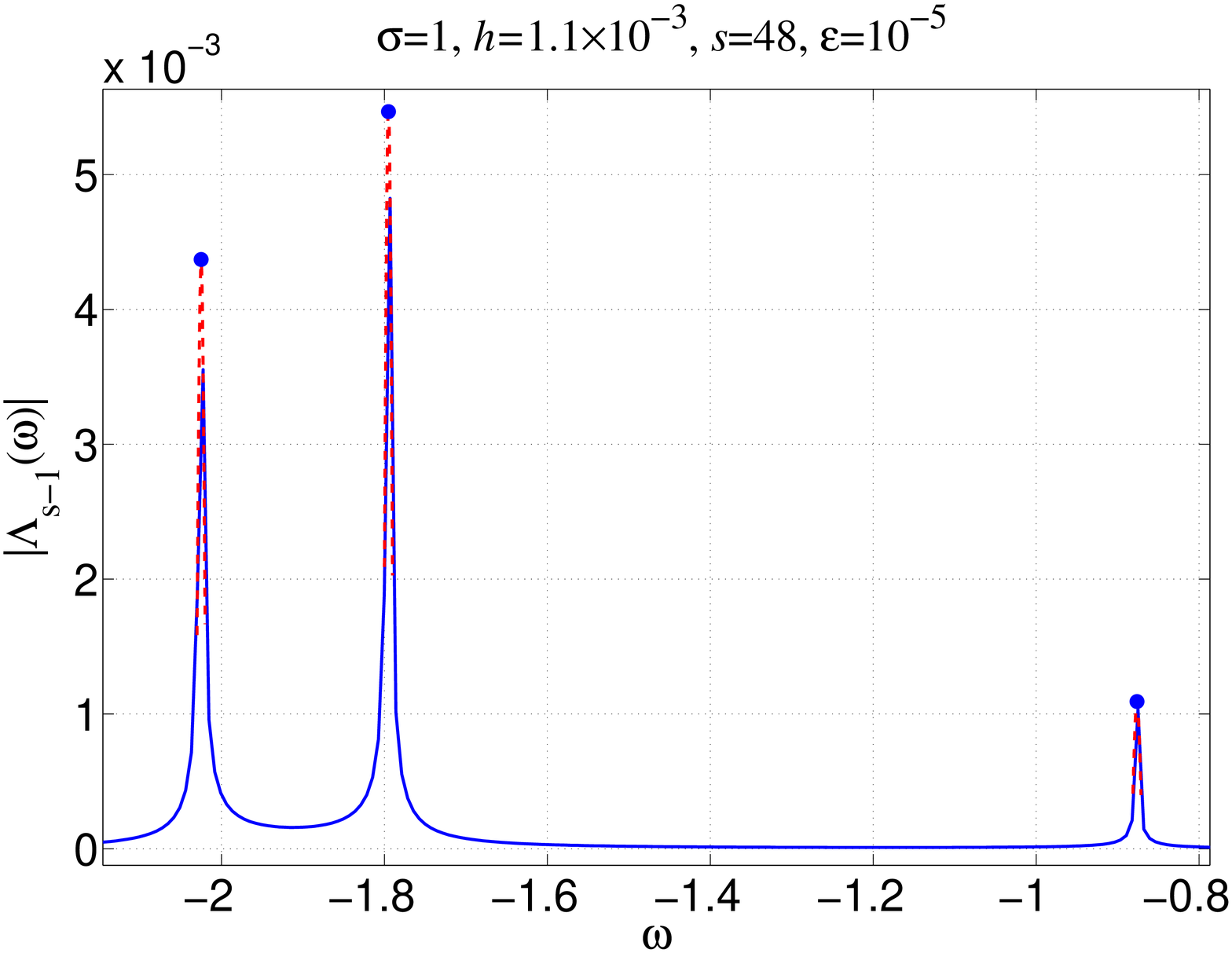}
\caption{\label{fig:epsart5} Numerically computed Fourier spectrum modulus ${\left| \Lambda_{s-1}^{(0)} \right|}$  (full and dashed curves) and (full circles) spectral amplitudes obtained from Eq. (\ref{ZeroOrderSm1}) in the stable region. Full curve: spectrum computed with the fast Fourier transform (FFT); dashed curves:  spectrum computed with the "slow" Fourier transform (SFT).}
\end{center}
\end{figure}

\begin{figure}
\begin{center}
\includegraphics[width=8.0cm]{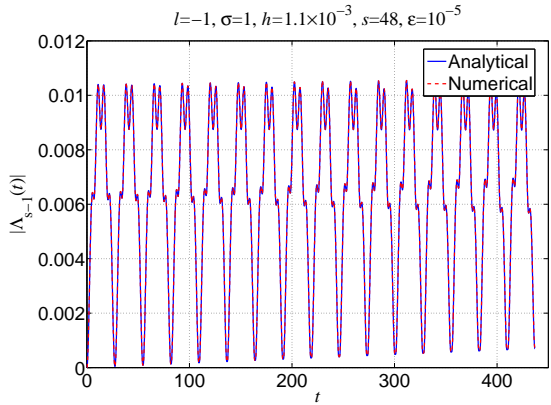}
\caption{\label{fig:epsart6} Time-domain numerical and analytical solutions for ${\left| \Lambda_{s-1}^{(0)} \right|}$  in the stable region.}
\end{center}
\end{figure}

A similar comparison has been performed in the stable region and the results are presented in both  frequency (Fig. \ref{fig:epsart5})  and time (Fig. \ref{fig:epsart6}) domains. A fast Fourier transform has been performed on $\Lambda_{s-1}^{(0)} = \Psi(s-1,t)$ and the result is presented by the full curve on Fig. \ref{fig:epsart5}. The three peaks correspond to the three real roots of Eq. (\ref{CharacteristicEq}). In order to increase the frequency resolution a slow Fourier transform has then been performed in the vicinity of the peaks (dashed lines in Fig. \ref{fig:epsart5}). Finally, the amplitudes of the three components of $\Lambda_{s-1}^{(0)}$ have been computed from Eq. (\ref{ZeroOrderSm1}) and plotted (the full circles in Figure \ref{fig:epsart5}). As can be seen the numerical and the analytical results are in excellent agreement.

\begin{figure}
\begin{center}
\includegraphics[width=8.0cm]{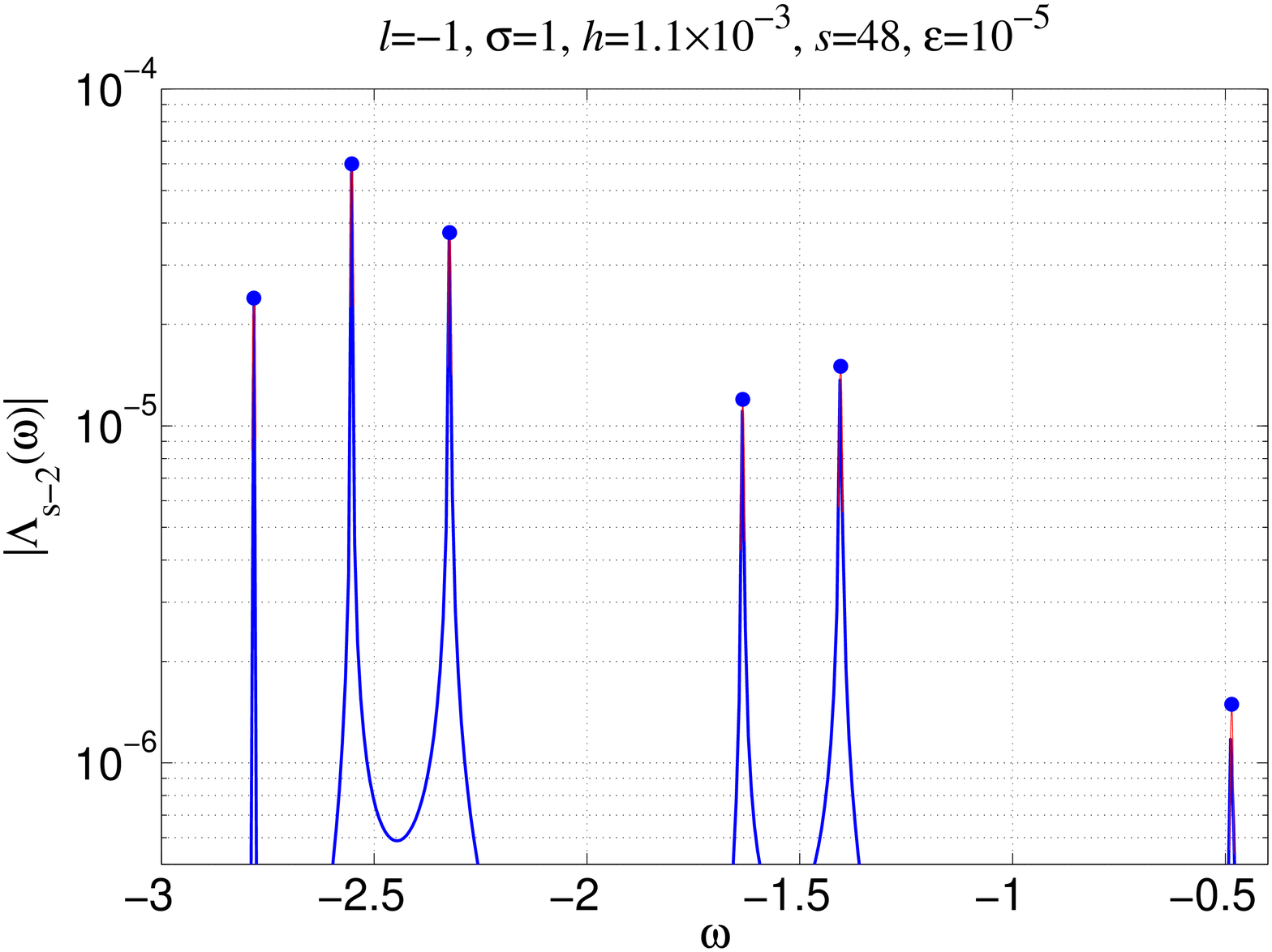}
\caption{\label{fig:epsart7} Frequency-domain solutions for ${\left| \Lambda_{s-2}^{(1)} \right|}$. The curves represent numerically computed spectra (by FFT and SFT) and the full circles are the spectral amplitudes obtained analytically from Eq. (\ref{FirstOrdSolLSm2}).}
\end{center}
\end{figure}

\begin{figure}
\begin{center}
\includegraphics[width=8.0cm]{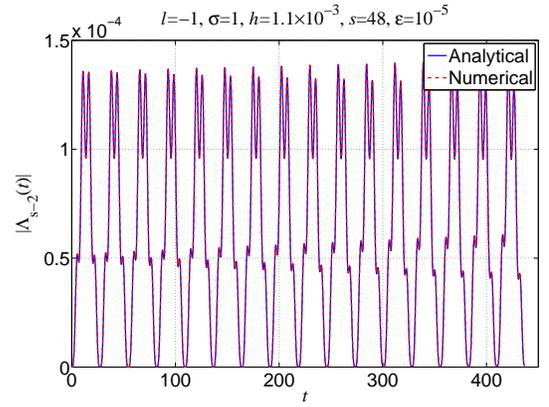}
\caption{\label{fig:epsart8} Time domain analytical and numerical solutions for ${\left| \Lambda_{s-2}^{(1)} \right|}$.}
\end{center}
\end{figure}

The same procedure has been applied to $\Lambda_{s-2}^{(1)} = \Psi(s-2,t)$ and the results are shown in Figures \ref{fig:epsart7} and \ref{fig:epsart8}. Note, that according to Eq. (\ref{FirstOrdSolLSm2}) the spectrum of $\Lambda_{s-2}^{(0)}$ contains a total of six discrete lines corresponding to combination frequencies of the type $\omega_{\beta} + \omega_{\gamma}$ in full agreement with what Fig. \ref{fig:epsart7} shows.

\begin{figure}
\begin{center}
\includegraphics[width=8.0cm]{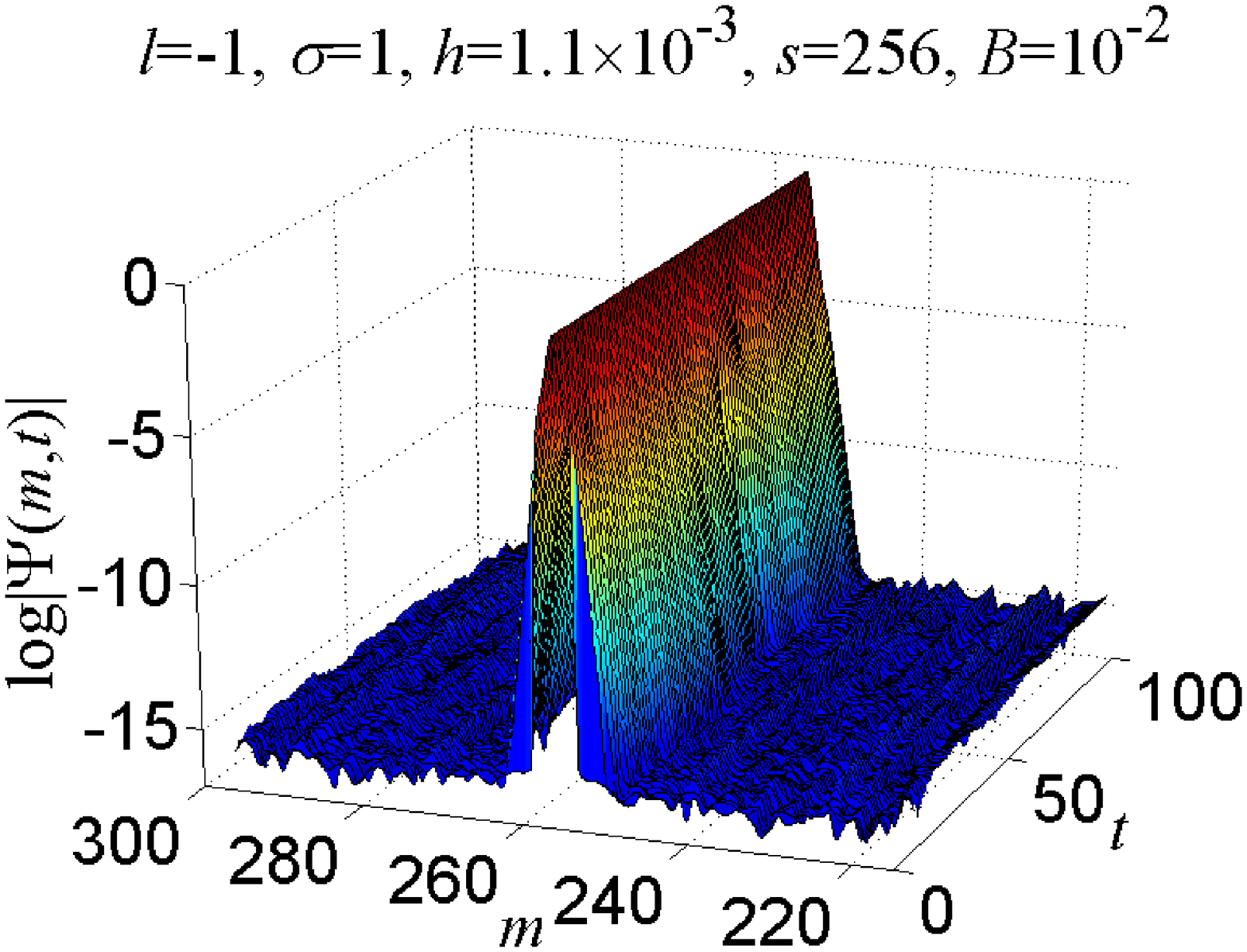}
\caption{\label{fig:epsart9} Solitary-wave-like behavior of ${\left| \Psi(m,t) \right|}$. The initial condition is given by Eqs. (\ref{FirstAStat}), (\ref{FirstBStat}) and (\ref{SecondSolStat}).}
\end{center}
\end{figure}

Figure \ref{fig:epsart9} shows the evolution of the modulus of the spectrum of the wave function ${\left| \Psi(m,t) \right|}$ obtained from the numerical solution of Eq. (\ref{Schrodinger}) with the amplitude of the single-wave mode $a$ given by Eq. (\ref{SingleWaveStation}). The initial condition $\Psi(m,t=0)$ is determined by Eqs. (\ref{FirstAStat}), (\ref{FirstBStat}) and (\ref{SecondSolStat}). As can be seen, a stable solitary-wave-like structure is formed.

\section{\label{sec:remarks}Concluding Remarks}

The present paper can be regarded as an extension of the classical single-wave model \cite{DelCastillo,MarTze} to the case, where quantum-mechanical effects begin to play comparable role. The quantum single-wave model has been derived starting from the von Neumann-Maxwell equations and closely following the basic steps outlined by del-Castillo-Negrete \cite{DelCastillo}. It is important to note that the basic equations in Schrodinger representation figuring as a starting point for our analysis can be obtained formally by canonical quantization of the classical single-wave model.

The linear stability of the quantum single-wave model has been studied in Sections \ref{sec:schrodpic} and \ref{sec:perturbative}. It has been shown that irrespective of the character of the spectrum of eigenfrequencies (real or complex), once a certain harmonic is generated, it excites its closest neighbors due to the self-consistent closest neighbor interaction law, which is intrinsic for the system. In the stable region (of parameter space, where all roots of the characteristic equation are real) periodic in time patterns are formed, which have been observed to be marginally stable. In the unstable region (one real and two complex conjugate roots of the characteristic equation) a numerically detected chain-type excitation of closest neighbors resembles a chaotic phenomenon. The freshly generated harmonics excite new ones, thus spreading gradually in harmonic number space until all the Fourier spectrum becomes populated.

Based on the method of multiple scales, a system of coupled nonlinear equations for the slowly varying amplitudes of interacting eigenmodes has been derived. As a matter of fact, the nonlinear standing-wave solutions of the single-wave model can be regarded as solutions of a nonlocal nonlinear Schrodinger equation.. A class of solutions has been found, which behave as a stable solitary-wave pattern.

\appendix

\section{\label{sec:appendix}Linear Stability Properties of the Quantum Single-Wave Model}

The von Neumann equation coupled with the equation for the single field mode has an evident equilibrium solution in the form $f = f_0 {\left( p \right)}$, $a = a_0 = 0$. To analyze its linear stability, we represent the Wigner quasi-probability distribution function as
\begin{equation}
f {\left( x, p; t \right)} = f_0 {\left( p \right)} + \sum \limits_{n = - \infty}^{\infty} f_n {\left( p; t \right)} e^{inx}, \label{AppWigner}
\end{equation}
and substitute it into Eq. (\ref{VonNeumannFin}). Retaining linear terms only, we obtain
\begin{equation}
\partial_t f_n + inp f_n = {\frac {i {\widetilde{a}} \delta_{n, \pm 1}} {\hbar}} {\left[ f_0 {\left( p - {\frac {\hbar} {2}} \right)} - f_0 {\left( p + {\frac {\hbar} {2}} \right)} \right]}, \label{AppLinNeumann}
\end{equation}
\begin{equation}
{\left( \sigma \partial_t + i l \right)} {\widetilde{a}} = i \int {\rm d} p f_1 {\left( p; t \right)}, \label{AppFieldMode}
\end{equation}
where ${\widetilde{a}}$ is a deviation from the equilibrium value $a_0 = 0$. We further assume that the linear modes evolve according to the law
\begin{equation}
f_n {\left( p; t \right)} = F_n {\left( p \right)} e^{- ict}, \qquad {\widetilde{a}} {\left( t \right)} = \alpha e^{- ict}, \label{AppLinModes}
\end{equation}
and readily obtain
\begin{equation}
F_n {\left( p \right)} = {\frac {\alpha \delta_{n, \pm 1}} {\hbar {\left( p - c \right)}}} {\left[ f_0 {\left( p - {\frac {\hbar} {2}} \right)} - f_0 {\left( p + {\frac {\hbar} {2}} \right)} \right]}. \label{AppLinDistrib}
\end{equation}
Finally, the dispersion relation can be expressed as
\begin{equation}
{\cal D} {\left( c \right)} = \sigma c - l - \int {\frac {{\rm d} p f_0 {\left( p \right)}} {{\left( p - c \right)}^2 - \hbar^2 / 4}} = 0. \label{AppDispRel}
\end{equation}
Let us take the equilibrium Wigner quasi-probability distribution in the form
\begin{equation}
f_0 {\left( p \right)} = {\frac {1} {2 \pi}} \delta {\left( p - \hbar s \right)}, \label{AppEquilWigner}
\end{equation}
which is consistent with the exact solution (\ref{ZeroOrderPsim}) of the Schrodinger equation (\ref{Schrodinger}). The dispersion relation (\ref{AppDispRel}) transforms then to the form
\begin{equation}
{\cal D} {\left( c \right)} = \sigma c - l - {\frac {1} {2 \pi}} {\frac {1} {{\left( \hbar s - c \right)}^2 - \hbar^2 / 4}} = 0. \label{AppDispRelat}
\end{equation}
Identifying now $c$ with $- \omega$, the latter expression coincides with the characteristic equation (\ref{CharacteristicEq}) obtained in Section \ref{sec:perturbative}.

\nocite{*}
\bibliography{aipsamp}

\end{document}